\documentclass[sigconf]{acmart}

\pdfoutput=1

\AtBeginDocument{%
  }

\usepackage{tikz}
\usepackage{pgfplots}
\usepackage{tikzlings}

\def \tfwidth{0.8\linewidth}
\def \tfheight{0.45\linewidth}

\usepackage{subcaption}

\usepackage{glossaries}
\newacronym[plural=IMUs,firstplural=Inertia Measurement Units]{imu}{IMU}{Inertial Measurement Unit}
\newacronym{vr}{VR}{Virtual Reality}
\newacronym{xr}{XR}{eXtended Reality}
\newacronym{hmd}{HMD}{Head-Mounted Display}
\newacronym{qoe}{QoE}{Quality of Experience}
\newacronym{upa}{UPA}{Uniform Planar Array}
\newacronym{hpbw}{HPBW}{half-power beam width}
\newacronym{aoa}{AoA}{Angle of Arrival}
\newacronym{aod}{AoD}{Angle of Departure}
\newacronym{toa}{ToA}{Time of Arrival}
\newacronym{csi}{CSI}{Channel State Information}
\newacronym[plural=TRNs,firstplural=Training Fields]{trn}{TRN}{Training Field}
\newacronym{music}{MUSIC}{MUltiple SIgnal Classification}
\newacronym{2dmusic}{2D-MUSIC}{2D MUltiple SIgnal Classification}
\newacronym{mimo}{MIMO}{Multiple Input Multiple Output}
\newacronym{simo}{SIMO}{Single Input Multiple Output}
\newacronym{cir}{CIR}{Channel Impulse Response}
\newacronym{cfr}{CFR}{Channel Frequency Response}
\newacronym{ula}{ULA}{Uniform Linear Array}
\newacronym{ue}{UE}{User Equipment}
\newacronym{bs}{BS}{Base Station}
\newacronym{tdd}{TDD}{Time Division Duplexing}
\newacronym{ofdm}{OFDM}{Orthogonal Frequency Division Multiplexing}
\newacronym{snr}{SNR}{Signal to Noise Ratio}
\newacronym{los}{LoS}{Line of Sight}
\newacronym{nlos}{NLoS}{Non Line of Sight}
\newacronym{tx}{TX}{Transmitter}
\newacronym{rx}{RX}{Receiver}
\newacronym{rf}{RF}{Radio Frequency}
\newacronym{mu}{MU}{Multi User}
\newacronym{mp}{MP}{Matrix Pencil}
\newacronym{rms}{RMS}{Root Mean Square}
\newacronym{fft}{FFT}{Fast Fourier Transform}
\newacronym{wlan}{WLAN}{wireless local area network}
\newacronym{ofdma}{OFDMA}{Orthogonal Frequency Division Multiplexing Access}
\newacronym{jcs}{JCAS}{Joint Communication And Sensing}
\newacronym{mmwave}{mmWave}{millimeter wave}
\newacronym{rmse}{RMSE}{Root Mean Square Error}
\newacronym{eirp}{EIRP}{equivalent isotropic radiated power}
\newacronym[plural=APs,firstplural=Access Points]{ap}{AP}{acces point}
\newacronym[plural=RISs,firstplural=Reconfigurable Intelligent Surfaces]{ris}{RIS}{Reconfigurable Intelligent Surface}
\newacronym[plural=PSDs,firstplural=Power Spectral Densities]{psd}{PSD}{Power Spectral Density}
\newacronym{ir}{IR}{InfraRed}
\newacronym{jcas}{JCAS}{Joint Communication and Sensing}

\usepackage{cleveref}
\usepackage{paralist}

\definecolor{mycolor1}{rgb}{0.00000,0,1}%
\definecolor{mycolor2}{rgb}{1,0,0}%
\definecolor{mycolor3}{rgb}{0,0,0}%
\definecolor{mycolor4}{rgb}{1,0,1}%

\copyrightyear{2022}
\acmYear{2022}
\setcopyright{acmcopyright}\acmConference[EmergingWireless '22 ]{Emerging Topics in Wireless }{December 9, 2022}{Roma, Italy}
\acmBooktitle{Emerging Topics in Wireless (EmergingWireless '22 ), December 9, 2022, Roma, Italy}
\acmPrice{15.00}
\acmDOI{10.1145/3565474.3569072}
\acmISBN{978-1-4503-9934-0/22/12}

\begin{document}

%%
%% The "title" command has an optional parameter,
%% allowing the author to define a "short title" to be used in page headers.
\title{DOPAMINE: Doppler frequency and Angle of arrival MINimization of tracking Error for extended reality}

\author{Andrea Bedin}
\affiliation{%
  \institution{Nokia Bell Labs}
  \country{Espoo, Finland}}
  \affiliation{%
  \institution{University of Padova, DEI}
  \country{Padova, Italy}}
\email{andrea.bedin.2@studenti.unipd.it}

\author{Alexander Marinšek}
\affiliation{%
    \institution{KU Leuven, ESAT-WaveCore}
    \country{9000 Ghent, Belgium}\\
}
\email{alexander.marinsek@kuleuven.be}

\author{Shaghayegh Shahcheraghi}
\affiliation{%
  \institution{Technical University of Darmstadt}
  \country{Darmstatd, Germany}}
\email{sshahcheraghi@wise.tu-darmstadt.de }

\author{Nairy Moghadas Gholian}
\affiliation{%
  \institution{Technical University of Darmstadt}
  \country{Darmstatd, Germany}}
\email{ngholian@wise.tu-darmstadt.de }

\author{Liesbet Van der Perre}
\affiliation{%
    \institution{KU Leuven, ESAT-WaveCore}
    \country{9000 Ghent, Belgium}\\
}
\email{liesbet.vanderperre@kuleuven.be}

\renewcommand{\shortauthors}{A. Bedin, A. Marinšek, S. Shahcheraghi, N. M. Gholian, and L. Van der Perre}

%%
%% The abstract is a short summary of the work to be presented in the
%% article.
\begin{abstract}
  In this paper, we investigate how \gls{jcs} can be used to improve the \gls{imu}-based tracking accuracy of \gls{xr} \glspl{hmd}. Such tracking is used when optical and \gls{ir} tracking is lost, and its lack of accuracy can lead to disruption of the user experience. In particular, we analyze the impact of using doppler-based speed estimation to aid the accelerometer-based position estimation, and \gls{aoa} estimation to aid the gyroscope-based orientation estimation. Although less accurate than \glspl{imu} for short times in fact, the \gls{jcs} based methods require one fewer integration step, making the tracking more sustainable over time. Based on the proposed model, we conclude that at least in the case of the position estimate, introducing \gls{jcs} can make long lasting optical/\gls{ir} tracking losses more sustainable.
\end{abstract}

%%
%% The code below is generated by the tool at http://dl.acm.org/ccs.cfm.
%% Please copy and paste the code instead of the example below.
%%



%%
%% Keywords. The author(s) should pick words that accurately describe
%% the work being presented. Separate the keywords with commas.


%%
%% This command processes the author and affiliation and title
%% information and builds the first part of the formatted document.
\maketitle

\glsresetall

\section{Introduction}

Accurately tracking the position of \gls{xr} \glspl{hmd} is critical for maintaining high user \gls{qoe}. Modern \glspl{hmd} achieve this using precise visible/\gls{ir} light tracking equipment. Unfortunately, visual/\gls{ir} tracking is also prone to outage in adverse circumstances. Although an \gls{hmd} can switch to \gls{imu} based tracking in such situations, long term visual/\gls{ir} outage will result in error buildup due to accelerometer and gyroscope drift integration.
\glspl{hmd} mostly rely on two major visual/\gls{ir} tracking techniques: a) the \gls{hmd} gathers visual/\gls{ir} information from an illuminated environment
or b) the \gls{hmd} receives visual/\gls{ir} signals from multiple external beacons in its vicinity. Examples of \glspl{hmd} fitted with a) and b) tracking options are the Oculus Quest 2 and the HTC Vive 2, respectively. Option a) is inherently prone to accuracy variations and outage due to limited control over environment illumination. Option b) offers precise tracking when the beacons lie within the \gls{hmd}'s field of view; yet, it experiences outage as soon as the \gls{hmd} looses sight of the external beacons \cite{niehorster_accuracy_2017}. A possible cause is either the user waving their hand or another object passing between a beacon and the \gls{hmd}. Moreover, the \gls{xr} users can rotate their head away from the IR light sources. Outage can befall \glspl{hmd} equipped with option a) tracking in a similar manner when the main light source is blocked. Hence, manufacturers should foresee visual/\gls{ir} tracking outage and provide the \gls{hmd} with alternative tracking solutions.
To avoid error-prone \gls{hmd}-based tracking during longer visual/\gls{ir} tracking outage, the wireless communication transceivers in an \gls{hmd} can double as a sensing device. The transceiver provides directly the \gls{hmd} orientation through \gls{aoa} estimation, while velocity is extracted from Doppler frequency shift estimation; thus, removing the need for one numerical integration step in both cases, compared to \gls{imu} tracking. Sensing the environment by estimating the \gls{aoa}, \gls{toa}, or Doppler frequency extracted from communication signals, called \gls*{jcas} has been proposed used in recent literature \cite{dokhanchi2019mmwave, kumari2019adaptive,robertieeead,kotaru2015spotfi}.
In the work at hand, we propose and evaluate a \textbf{Dop}pler frequency and \textbf{A}oA technique for \textbf{min}imizing \gls{xr} \gls{hmd} tracking \textbf{e}rror during visual/\gls{ir} tracking outage, called Dopamine. The proposed approach can be used alongside \gls{imu}-based tracking and substitute it dynamically upon error accumulation. Dopamine is primarily designed to operate in the 60~GHz \gls{mmwave} frequency band and loosely conform with conventions standardized in IEEE~802.11ad/ay. Moreover, Dopamine's flexibility allows for easy portability to other frequency bands and standards. Our contributions are three-fold: I) the extraction of \gls{hmd} tracking data using limited wireless channel information, II) the analysis of tracking performance when employing either multiple \glspl{ap} or multiple \glspl{ris}, and III) the evaluation of the proposed tracking technique against \gls{imu}-based tracking.

The paper is structured as follows: \Cref{sec:system_model} defines a simple scenario that will be used for the evaluation, \Cref{sec:estimation} describes the Dopamine tracking and estimation framework, \Cref{sec:results} presents and discusses the performance results, while \Cref{sec:conclusion} draws a conclusion on the study's findings and highlights future research areas.

\section{System model}
\label{sec:system_model}

The following subsections describe the considered scenario, including the environment description, wireless signal propagation modeling, and an overview of \gls{xr} user mobility. \Cref{tab:system_parameters} lists the most relevant system parameters.

\begin{table*}
  \caption{System parameters.}
  \vspace{-0.15cm}
  \label{tab:system_parameters}
  \begin{tabular}{cc|cc}
    \toprule
    Name&Value&Name&Value\\
    \midrule
    
    % Environment
    
    Room dimensions & 4x4x2.8~m & HMD\,/\,AP\,/\,RIS height & 1.8~m \\
    AP$_1$-RIS$_2$\,/\, distance & 2.8~m & AP$_1$\,/\,AP$_2$\,/\,RIS$_1$-HMD distance & 2~m \\

    \midrule
    
    % RF
    
    Carrier frequency ($f_{0}$) & $60$~GHz & Transmit power ($P_{TX}$) & 24~dBm \\
    Thermal noise power & -82~dBm & Noise figure & 10~dB \\
    Receive gain ($G_{RX}$) &$14$--$17$~dB & HMD antenna HPBW & 120$^\circ$  \\
    HMD antenna arrays & 4 & HMD array size ($M \times N$) & 4x4 \\
    RIS gain ($G_{RIS}$) & 0~dB & x & x \\

    \midrule
    
    % Application
    
    \gls{hmd} display refresh rate & 120~Hz  & PHY packet rate (R$_p$) & 50 Kpkt/s \\
    Movement bandwidth ($B_m$) & 30~Hz & Beam training interval ($T_{b}$) & 102.4ms \\

  \bottomrule
\end{tabular}
\end{table*}

\subsection{Environment}

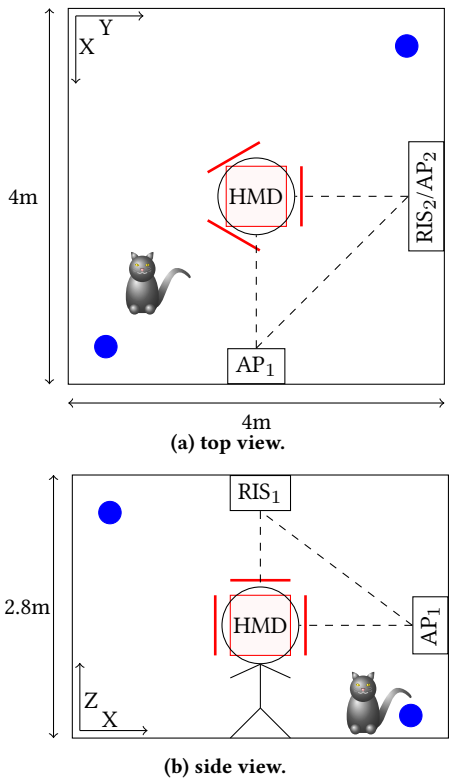
\begin{figure}[t]
     \centering
     \begin{subfigure}[b]{0.35\textwidth}
         \centering
         \begin{tikzpicture}

\def\size{5}
\def\center{\size/2}
\def\arraysize{0.4}
\def\arrayoffset{0.6}

\draw[color=red, fill=red, fill opacity=0.03] (\center-\arraysize,\center-\arraysize) -- (\center+\arraysize,\center-\arraysize) -- (\center+\arraysize,\center+\arraysize) -- (\center-\arraysize,\center+\arraysize) -- cycle;
%\draw[color=red,line width=0.35mm] (\center+\arrayoffset,\center-\arraysize) -- (\center+\arrayoffset,\center+\arraysize);
%\draw[color=red,line width=0.35mm] (\center-\arrayoffset,\center-\arraysize) -- (\center-\arrayoffset,\center+\arraysize);
%\draw[color=red,line width=0.35mm] (\center-\arraysize,\center+\arrayoffset) -- (\center+\arraysize,\center+\arrayoffset);
\draw[color=red,line width=0.35mm] (\center - 0.5*\arrayoffset+0.866*\arraysize,\center+0.866*\arrayoffset+0.5*\arraysize) -- (\center - 0.5*\arrayoffset-0.866*\arraysize,\center+0.866*\arrayoffset-0.5*\arraysize);
\draw[color=red,line width=0.35mm] (\center - 0.5*\arrayoffset+0.866*\arraysize,\center-0.866*\arrayoffset-0.5*\arraysize) -- (\center - 0.5*\arrayoffset-0.866*\arraysize,\center-0.866*\arrayoffset+0.5*\arraysize);
\draw[color=red,line width=0.35mm] (\center+\arrayoffset,\center-\arraysize) -- (\center+\arrayoffset,\center+\arraysize);

\draw (0,0) -- (\size,0) -- (\size,\size) -- (0, \size) -- cycle;
\node[draw, circle, color=black] (h) at (\center, \center) {HMD};
\node[draw, rectangle, color=black,  anchor=south] (ap) at (\center, 0) {AP$_1$};
\node[draw, rectangle, color=black,  anchor=south, rotate=90] (ris) at (\size, \center) {RIS$_2$/AP$_2$};

\draw[dashed] (ap.north) -- (h.south);
\draw[dashed] (ap.north) -- (ris.north);
\draw[dashed] (ris.north) -- (h.east);

\draw[<->] (0, -0.25) -- (\size, -0.25); 
\node at (\size/2, -0.5) {4m};

\draw[->] (0.1,\size - 0.1) -- (1, \size -0.1);
\draw[->] (0.1, \size -0.1) -- (0.1, \size -1);

\node at (0.25, \size -0.5) {X};
\node at (0.5, \size -0.25) {Y};

\draw[<->] ( -0.25, 0) -- (-0.25,\size); 
\node at (-0.6, \size/2) {4m};

% \draw[color=gray] (1-\arraysize,1-\arraysize) -- (1+\arraysize,1-\arraysize) -- (1+\arraysize,1+\arraysize) -- (1-\arraysize,1+\arraysize) -- cycle;

%\node [draw, color=blue] at (0.5, 0.5) {IR$_1$};
%\node [draw, color=blue] at (\size-0.5, \size-0.5) {IR$_2$};

\node [draw, circle, color=blue, fill=blue] at (0.5, 0.5) {};
\node [draw, circle, color=blue, fill=blue] at (\size-0.5, \size-0.5) {};

%\node [draw, circle, color=gray] at (1.2, 1.2) {BL};
\cat[body=gray,xshift=28,yshift=25,scale=0.4, 3D] at (1.2, 1.2);

\end{tikzpicture}
         \vspace{-0.3cm}
         \caption{top view.}
         \label{fig:topview}
     \end{subfigure}
    
     \begin{subfigure}[b]{0.35\textwidth}         
         \centering
         \vspace{0.2cm}
         \begin{tikzpicture}

\def\size{5}
\def\center{\size/2}
\def\height{3.5}
\def\headhegiht{1.5}
\def\arraysize{0.4}
\def\arrayoffset{0.6}

\draw (0,0) -- (\size,0) -- (\size,\height) -- (0, \height) -- cycle;

\draw[color=red, fill=red, fill opacity=0.03] (\center-\arraysize,\headhegiht-\arraysize) -- (\center+\arraysize,\headhegiht-\arraysize) -- (\center+\arraysize,\headhegiht+\arraysize) -- (\center-\arraysize,\headhegiht+\arraysize) -- cycle;
\draw[color=red,line width=0.35mm] (\center+\arrayoffset,\headhegiht-\arraysize) -- (\center+\arrayoffset,\headhegiht+\arraysize);
\draw[color=red,line width=0.35mm] (\center-\arrayoffset,\headhegiht-\arraysize) -- (\center-\arrayoffset,\headhegiht+\arraysize);
\draw[color=red,line width=0.35mm] (\center-\arraysize,\headhegiht+\arrayoffset) -- (\center+\arraysize,\headhegiht+\arrayoffset);

\node[draw, circle, color=black] (h) at (\center, \headhegiht) {HMD};

\node[draw, rectangle, color=black,  anchor=south, rotate=90] (ap) at (\size, \headhegiht) {AP$_1$};
% \node[draw, rectangle, color=black,  anchor=north] (ris) at (\center, \height) {RIS/AP};
\node[draw, rectangle, color=black,  anchor=north] (ris) at (\center, \height) {RIS$_1$};

\draw[dashed] (ap.north) -- (h.east);
\draw[dashed] (ap.north) -- (ris.south);
\draw[dashed] (ris.south) -- (h.north);

\draw (h.south) -- (\center, 0.4) ;

\draw (h.south) -- (\center-0.4, 0.8) ;
\draw (h.south) -- (\center+0.4, 0.8) ;

\draw (\center - 0.4, 0)  -- (\center, 0.4) ;
\draw (\center + 0.4, 0)  -- (\center, 0.4) ;

\draw[->] (0.1, 0.1) -- (1, 0.1);
\draw[->] (0.1, 0.1) -- (0.1, 1);

\node at (0.25, 0.5) {Z};
\node at (0.5, 0.25) {X};

\draw[<->] ( -0.25, 0) -- (-0.25,\height); 
\node at (-0.6, \height/2) {2.8m};

\node [draw, circle, color=blue, fill=blue] at (\size - 0.5, 0.3){};
\node [draw, circle, color=blue, fill=blue] at (0.5, \height-0.5){};

\cat[body=gray,xshift=110,scale=0.4, 3D];
\end{tikzpicture}
         \caption{side view.}
         \label{fig:sideview}
     \end{subfigure}
     \vspace{-0.2cm}
        \caption{Considered scenario. Dashed black lines represent signal propagation paths between the \glspl{ap}, \glspl{ris}, and the \gls{hmd}. UPAs are marked in red, while the two \gls{ir} light tracking beacons are depicted in blue and the blocker is depicted in gray.} \vspace{-5pt}
        \label{fig:environment}
\end{figure}

\Cref{fig:environment} depicts the considered environment. Although it is very specific, we  consider it is representative of a realistic \gls{xr} scenario and the results obtained will be indicative of the potential improvement in tracking brought by the method. An \gls{xr} user is positioned in the middle of a square room with a 2.8~m high ceiling and measuring 4~m in length and width. The main \gls{ap} (AP$_1$) is placed on the southern wall at a height of 1.8~m, while the main \gls{ris} (RIS$_1$) is attached to the ceiling, directly above the user. Either a second \gls{ap} or \gls{ris} (AP$_2$/RIS$_2$) is placed on the eastern wall at the same height as AP$_1$. The use of \glspl{ris} is essential to ensure that the system has a reasonable cost: mmWave \glspl{ap} are very costly, and multi-\gls{ap} solutions might not become available in the near future.

The \gls{hmd} is equipped with 4 square \glspl{upa}. One is attached to the the top headband of the \gls{hmd}, orienting it directly towards RIS$_1$. The remaining 3 are spread out along the \gls{hmd}'s frame and horizontal headband, misaligned by 120$^{\circ}$. This provides full azimuth coverage since the considered microstrip antennas with 5~dBi directivity have a 120$^{\circ}$ \gls{hpbw} \cite{yang_broadening_2018}. Thus, the \gls{hmd} receiver design features 4 RF chains, one for each \gls{upa}, of which only 3 are employed for tracking purposes at any one time. Compared to a fully digital MIMO system, this is more feasible in resource-constrained consumer devices, which we assume will anyway feature multiple phased antenna arrays for coping with mobility and adverse circumstances.

Two additional \gls{ir} light tracking beacons are foreseen in the room to highlight the problem at hand. The beacons are placed in two corners of the room, and we assume one is covered by an obstructing object. The wireless data link can serve as a supportive tracking technology as long as the visual/\gls{ir} beacons and the \glspl{ap}/\glspl{ris} are not co-located.

\subsection{Wireless signal propagation}

Determining the received signal strength at the \gls{hmd} $P_{RX}$ is the basis for evaluating Dopamine's performance as an RF-tracking alternative. Without further denoting the dependency on system parameters for now, the received signal strength is derived as follows:
\begin{equation}
    P_{RX} = P_{TX} + G_{RX} + G_{RIS} - PL
\end{equation}
\noindent where $P_{TX}$ is the transmitted signal power -- set to 24~dBm as in TP-Link Talon AD7200 \glspl{ap} \cite{talon_eirp} -- $G_{RX}$ and $G_{RIS}$ are the ensemble \gls{rx} and \gls{ris} power gain, respectively, while $PL$ corresponds to signal attenuation due to path loss.

\gls{rx} power gain is governed by antenna element directivity and array gain. The 120$^\circ$ \gls{hpbw} antenna radiation patterns are modeled as a cosine of the azimuth angle~\cite{itu_m18151}, while the number of antenna elements contributes to the array gain according to:
\begin{equation}
    G_{RX}(\varphi, M) = \underbrace{10\,log_{10}\,(\pi cos(\varphi))}_{antenna\;gain} + \underbrace{10\,log_{10}\,(M N)}_{array\;gain}
\end{equation}
where $\varphi$ is the impinging wave's azimuth angle and $M = N = 4$ represents the number of array elements, in the vertical and horizontal direction. We consider the worst case azimuth misalignment $\pm60^\circ$ for the three \glspl{upa} in the horizontal plane in all further calculations. This may happen while the \gls{xr} user is rotating. The misalignment reduces antenna gain to 2~dBi, yielding a 14~dB receive gain. Contrarily, the \gls{upa} on the top headband has boresight conditions to RIS$_1$, resulting in 17~dB gain. 
We apply the \gls{ris} power radiation pattern from \cite{RIS_gain} in both receive and transmit direction due to reciprocity. We assume the \glspl{ris} are sufficiently small (e.g. a 5x5~cm panel with 10x10 elements), resulting in a far-field distance less than 1~m.The ceiling-mounted RIS$_1$ is oriented so that it's 0-elevation plane crosses through both the \gls{hmd} and AP$_1$, giving it unitary gain. Similarly, $G_{RIS}=0$~dB applies to RIS$_2$ if its 0-elevation plane is level to the ground.
Attenuation due to the propagation distance is modelled according to the free-space path loss equation: $PL = 20\,log_{10}\, \left( \frac{4 \pi d}{\lambda} \right)$,  where $d$ represents the total length of the propagation path and $\lambda$ is the 5~mm wavelength at 60~GHz.
% Finally, the \gls{snr} at the \gls{hmd} is derived using:
Finally, we use the derived receive power $P_{RX}$ to calculate the \gls{snr}: $SNR = P_{RX} - P_N - NF$, where $P_N$ corresponds to thermal noise at room temperature (300~K) for the IEEE 802.11ad/ay 1.76~GHz channel bandwidth (without guard frequencies) and $NF$ stands for an additional 10~dB noise figure \cite{ieee_80211-2020_2021}. \Cref{tab:worst_case_snr_per_path} lists the derived \gls{snr} values for each of the considered propagation paths. Note, that the first two values are equal since the 3~dB loss due to user yaw misalignment (60$^\circ$) is identical to the added path loss in AP$_1$--RIS$_1$--HMD. Beam training sequences (e.g. BRP) can be appended to PHY packets with minimal overhead in case of low gain when beaming in multiple directions, i.e. towards RIS$_1$ and RIS$_2$.

\begin{table}[t]
 \centering
  \caption{Worst-case \gls{snr} for each scenario.}
  \vspace{-0.15cm}
  \label{tab:worst_case_snr_per_path}
  \begin{tabular}{cc}
    \toprule
    Path & SNR [dB]\\
    \midrule
    AP$_1$--HMD or AP$_2$--HMD & 36 \\
    AP$_1$--RIS$_1$--HMD &  36 \\
    AP$_1$--RIS$_2$--HMD &  28 \\
  \bottomrule
\end{tabular}
\end{table}

\subsection{XR user mobility overview} \label{sec:movement}

\begin{figure}[t]
     \centering
         \input{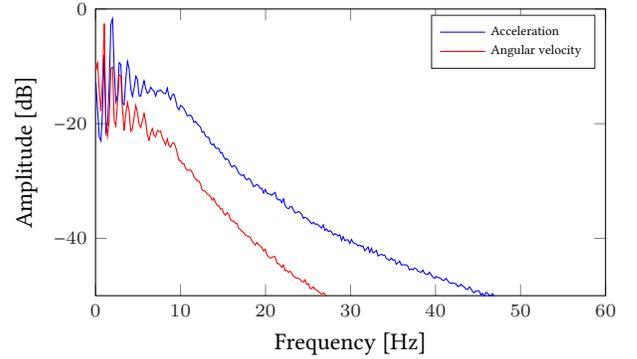}
         \vspace{-0.35cm}
        \caption{Normalized spectrum of the acceleration and angular velocity.}
          \vspace{-0.2cm}
        \label{fig:acc_spec}
\end{figure}

Understanding the characteristics of the user's mobility is necessary to properly design a tracking system. For this purpose, we used the dataset from \cite{RIDI} to determine the frequency content of the movements. 
In particular, we selected all the files reporting acceleration data and gyroscope data, estimated the \gls{psd} of the signal using Welch's algorithm for each file and averaged all the \glspl{psd} over all files. In the result, shown in figure \ref{fig:acc_spec}, we observe that for frequency over $B_m = 30$Hz the normalized \gls{psd} has negligible amplitude. For this reason, we require the sensors to be sampled more than $30$ times per second. Moreover, we also evaluated the \gls{rms} and maximum value for the acceleration and angular velocity thought the whole dataset. The resulting values are of $38^\circ/s$ and $261^\circ/s$ for the gyroscope and $1.8 \frac{\text{m}}{\text{s}^2}$  and $14 \frac{\text{m}}{\text{s}^2}$ for the accelerometer.

\section{Estimation}
\label{sec:estimation}

In this section, we describe how the position and rotation of the headset can be tracked using \gls{jcs}, and compare the radio-based tracking with the \gls{imu}-based tracking. 

\subsection{MUSIC} \label{sec:music}
In this paper, we use \gls{music} to perform spectral and \gls{aoa} estimations. \gls*{music} is a subspace-based high resolution algorithm which divides the received signal space into two subspaces, called the signal subspace and the noise subspace. It relies on the eigenvalue decomposition of the covariance matrix of the received signal, $\boldsymbol{R}_y=\boldsymbol{E}[\boldsymbol{y}\boldsymbol{y}^H]$, where $\boldsymbol{y}$ is the signal of interest impaired by noise. \gls*{music} computes all the eigenvalues and their associated eigenvectors of $\boldsymbol{R}_y$, and it splits the eigenvalues into two categories based on their magnitude. It considers the eigenvectors corresponding to the "large" eigenvalues as the vectors that span the signal subspace and the eigenvectors corresponding to the "near-zero" eigenvalues to span the Noise-subspace. The latter are the columns of the  so-called noise matrix $\boldsymbol{U}_n$. The estimated frequency or \gls{aoa} of the incident signals are the peaks of  \gls*{music} spectrum $P_{MUSIC}(X)=\frac{1}{\boldsymbol{a}^H(X)\boldsymbol{U}_n \boldsymbol{U}^H_n \boldsymbol{a}(X)}$, where $\boldsymbol{U}_n $ is the noise matrix, and $\boldsymbol{a}(X)$ is the steering vector associated with $X$, and $X$ is either a frequency, in case of spectral estimation, a pair of angles in case of 2D \gls{aoa} estimation or a single angle in case of 1D \gls{aoa} estimation.

\subsection{Velocity}
\begin{figure}[t]
     \centering
         % This file was created by matlab2tikz.
%
%The latest updates can be retrieved from
%  http://www.mathworks.com/matlabcentral/fileexchange/22022-matlab2tikz-matlab2tikz
%where you can also make suggestions and rate matlab2tikz.
%

\begin{tikzpicture}

\begin{axis}[%
xticklabel style={font=\footnotesize\color{white!15!black}},
yticklabel style={font=\footnotesize\color{white!15!black}},
xlabel near ticks,
ylabel near ticks,
width=\tfwidth,
height=\tfheight,
scale only axis,
xmode=log,
xmin=1,
xmax=50,
xminorticks=true,
xlabel style={font=\color{white!15!black}},
xlabel={$T$ [ms]},
ymin=0,
ymax=0.02,
scaled y ticks=false,
ylabel style={font=\color{white!15!black}},
ylabel={RMSE [m/s]},
yticklabel style={
        /pgf/number format/fixed,
        /pgf/number format/precision=5
},
axis background/.style={fill=white},
legend style={legend cell align=left, align=left, draw=white!15!black, font=\tiny, row sep=-1.5pt}
]
\addplot [color=mycolor1]
  table[row sep=crcr]{%
1	0.0180145822724677\\
1.93069772888325	0.0123015654447439\\
3.72759372031494	0.00527520628201849\\
7.19685673001152	0.00248660156243753\\
13.8949549437314	0.00136701766378664\\
26.8269579527973	0.000633167254986662\\
51.7947467923121	0.000343222443851464\\
};
\addlegendentry{$\bar{\sigma}_a = \text{0m/s}^\text{2}$}

\addplot [color=mycolor2]
  table[row sep=crcr]{%
1	0.0174146777030747\\
1.93069772888325	0.0120149300161698\\
3.72759372031494	0.00562642247602054\\
7.19685673001152	0.00341731806702621\\
13.8949549437314	0.00256370589009687\\
26.8269579527973	0.00264874725379216\\
51.7947467923121	0.00219174653534615\\
};
\addlegendentry{$\bar{\sigma}_a  = \text{10m/s}^\text{2}$}

\addplot [color=mycolor3]
  table[row sep=crcr]{%
1	0.0169407802471231\\
1.93069772888325	0.0125700498905171\\
3.72759372031494	0.00668433420125215\\
7.19685673001152	0.00553395978882058\\
13.8949549437314	0.00526303447242487\\
26.8269579527973	0.00513748191984371\\
51.7947467923121	0.00478344558830108\\
};
\addlegendentry{$\bar{\sigma}_a  = \text{20m/s}^\text{2}$}

\addplot [color=mycolor4]
  table[row sep=crcr]{%
1	0.0208741088470666\\
1.93069772888325	0.0155808551447578\\
3.72759372031494	0.0116582856112863\\
7.19685673001152	0.010668294849795\\
13.8949549437314	0.00979034747755678\\
26.8269579527973	0.00968693336784883\\
51.7947467923121	0.0102582234098236\\
};
\addlegendentry{$\bar{\sigma}_a  = \text{40m/s}^\text{2}$}

\end{axis}

\end{tikzpicture}%
         \vspace{-0.35cm}
        \caption{Velocity \gls{rmse} vs observation time.}\vspace{-0.2cm}
        \label{fig:mse_speed_vs_obs_time}
\end{figure}
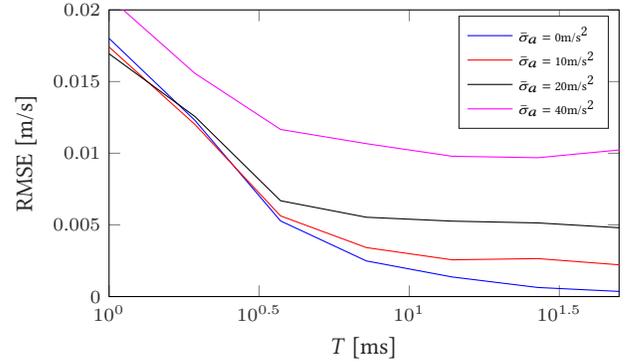

The velocity of the headset can be estimated based on the doppler effect measured on the received signal. This choice is mainly driven by the characteristics  of 802.11ad/ay: as the Doppler frequency is proportional to the carrier frequency,  mmWave systems can be very sensitive to movements provide very accurate velocity estimation. In particular, for the $k$-th received packet we assume to obtain a \gls{cfr} estimate from a beam pointing in the $x$, $y$ and $z$ direction, and that the received signal is dominated by the \gls{los} to the $\text{AP}_1$,  $\text{AP}_2$/$\text{RIS}_2$ and $\text{RIS}_1$ respectively. In the following, we only consider the movement in the $x$ direction, as the other directions can be modelled in the same way. We also assume that the initial position of the device is at $0$, i.e. at the beginning of the outage the receiver is perfectly synchronized (i.e. at time $0$ the frequency response is constant). With this assumption, the channel \gls{cfr} can be written as
\begin{equation}
    H(k,f) = \alpha e^{- \frac{j 2 \pi (f_0 + f) \sum_{i=0}^{k} v_x(i) }{c R_p}}
\end{equation}
Where $\alpha$ is assumed to be a constant channel coefficient and $v_x(i)$ is the velocity of the headset along the $x$ direction. In order to observe the effect of movement, we assume to observe the channel $K$ times. If the speed and orientation are constant throughout the observation, i.e. $v_X(i) = v_x$ for some constant $v_x$, and using the approximation $(f_0 + f) \approx f_0$,  since $f << f_0$, we can approximate the channel to:
\begin{equation}
    H(k,f) = \alpha e^{- \frac{j 2 \pi f_0 k v_x }{c R_p}} = \alpha e^{-j 2 \pi f_d k}.
\end{equation}
Where $ f_{d} = \frac{f_0 v_x }{c R_p}$ is the frequency of the channel variation over frame. Such frequency can be estimated using the \gls{music} algorithm described in section \ref{sec:music}. The estimated velocity can therefore be computed as $\bar{v}_x = v_x +  \mathcal{N}(0, \sigma_{v_x}) = \frac{c f_d R_p}{f_0}$.
%\begin{equation}
%    \bar{v}_x = v_x +  \mathcal{N}(0, \sigma_{v_x}) = \frac{c f_d R_p}{f_0}.
%\end{equation}
The observation period length $T = K R_p$ must be designed with the movement characteristics in mind: This value is subject to an inherent trade-off, as a low value gives worse frequency estimates, but for high values the assumption of constant velocity might not be satisfied. In this work, we choose to determine this value empirically by simulating the estimation for different values of $T$. For each value, we generated $100$ channel realizations with an \gls{snr} of $30$dB, we used a time varying speed $v_x(k) = \frac{k a}{R_p}$, where the acceleration $a$ is a realization of a normal distribution with zero mean and standard deviation $\bar{\sigma}_a$. We repeated the experiment for $\bar{\sigma}_a  \in \{0, 10, 20, 40 \} \frac{\text{m}}{\text{s}^2}$ and compared the result with the average velocity over the observation $\bar{v} = \frac{1}{K}\sum_{i=0}^{K-1}\frac{i a}{R_p}$. In Figure \ref{fig:mse_speed_vs_obs_time} we can see the \gls{rmse} of the estimated velocity as a function of the observation time $T$. Here we can observe how, for the considered values of acceleration and observation period, increasing the observation period always improves the estimate. For this reason, we decide to base the choice of $T$ on the \gls{hmd} display refresh rate instead, and use $T = \frac{1}{120Hz} = 8.3ms$. Indeed, we can check that this hypothesis leads to a small maximum angle deviation of $261^\circ/s \cdot T = 2.2^\circ$ over the observation. In order to have a fair comparison, we assume to use the same sampling rate also for the \gls{imu}.
With these assumptions, the \gls{rmse} of the velocity for the three components can be estimated numerically. In order to observe the effect of different movement patterns on the system, we evaluate the error for two acceleration profiles: a)  Profile P1: $\bar{\sigma}_a = 10 \frac{\text{m}}{\text{s}^2}$. b) Profile P2: $\bar{\sigma}_a = 40 \frac{\text{m}}{\text{s}^2}$.
The corresponding values for each propagation path and profile are listed in table  \ref{tab:speed_PSD}.

\begin{table}
  \caption{Speed \gls{rmse} $ \sigma_v \left( \frac{mm}{s}\right) $.}
  \label{tab:speed_PSD}
  \begin{tabular}{cccc}
    \toprule
    Path & profile 1  & profile 2\\
    \midrule
    AP$_1$--HMD or AP$_2$--HMD &  9.8 & 3  \\
    AP$_1$--RIS$_1$--HMD &  9.8 & 3\\
    AP$_1$--RIS$_2$--HMD & 10.9 & 4.4 \\
  \bottomrule
\end{tabular}
\end{table}

\subsection{Position}
 We now estimate the velocity for a sequence of $L$ observations. For the $\ell$-th observation, calling the noisy estimated velocity $\bar{v}_x^{(\ell)} = v_x^{(\ell)} + \mathcal{N}(0, \sigma_{v_x})$, the position estimate is obtained by integrating the $\bar{v}_x^{(\ell)}$ with respect to $\ell$.
The positional error is therefore the integral of a white Gaussian noise with a standard deviation of $T \sigma_x$, which results in a standard deviation of the position $  \sigma_{p_x}(t) = T \sigma_{v_x} \sqrt{\frac{t}{T}} = \left( \sqrt{T} \sigma_{v_x} \right) \sqrt{t}$.
%\begin{equation}
%    \sigma_{p_x}(t) = T \sigma_{v_x} \sqrt{\frac{t}{T}} = \left( \sqrt{T} \sigma_{v_x} \right) \sqrt{t}.
%\end{equation}
Assuming uncorrelated errors in the three components of the vector, the expected \gls{rms} amplitude of the 3D positional noise is therefore $\Vert \epsilon_p \Vert_{RMS}= \sqrt{(\sigma_{v_x}^2 + \sigma_{v_y}^2 + \sigma_{v_z}^2) T t}$.\\
%\begin{equation}
%  \Vert \epsilon_p \Vert_{RMS}= \sqrt{(\sigma_{v_x}^2 + \sigma_{v_y}^2 + \sigma_{v_z}^2) T t}.
%\end{equation}
By contrast, an accelerometer with an \gls{rms} noise of $\sigma_a$ would result in an acceleration noise process with standard deviation $T \sigma_a$ that, when doubly integrated, leads to an overall positional \gls{rmse} of 
\begin{equation}
    \Vert \epsilon_p \Vert_{RMS} = \sqrt{3 * \left(\frac{T \sigma_a}{\sqrt{3}} \left( \frac{t}{T} \right)^{\frac{3}{2}} \right)^2} = \frac{\sigma_{a} t ^{\frac{3}{2}}}{T}
\end{equation}
assuming uncorrelated noise in the three components.

\subsection{Angle}
Unlike speed estimation, the \gls{aoa} estimation requires full \gls{csi} for every antenna, so it can only be performed when the beam training is executed. In this paper, we assume that the receiver uses a codebook of $16$ orthonormal weights vectors, and call $\mathbf{B}$ the unitary matrix that has such vectors as rows. Calling $H_{i}(f)$ the channel observed by antenna $i$, we have that the vector of channel $\hat{H}_{k}(f)$ measured for each beam $k \in [1, 16]$ is
\begin{equation}
\left[\begin{matrix}
\hat{H}_{1}(f) \\
\vdots \\
\hat{H}_{16}(f)
\end{matrix}  \right] =\mathbf{B} \left(
\left[\begin{matrix}
H_{1}(f) \\
\vdots \\
H_{16}(f)
\end{matrix}  \right] + \mathcal{N}(0, \sigma  I)\right),
\end{equation}
from which we obtain the noisy estimate of the channel for each antenna
\begin{equation}
\left[\begin{matrix}
\bar{H}_{1}(f) \\
\vdots \\
\bar{H}_{16}(f)
\end{matrix}  \right] = 
\mathbf{B}^H \left[\begin{matrix}
\hat{H}_{1}(f) \\
\vdots \\
\hat{H}_{16}(f)
\end{matrix}  \right] =
\left[\begin{matrix}
H_{1}(f) \\
\vdots \\
H_{16}(f)
\end{matrix}  \right] + \mathcal{N}(0, \sigma I).
\end{equation}
Denoting the elevation and azimuth angle by $\theta$ and $\phi$ respectively, we apply the \gls*{music} algorithm to the estimated channel for the signal received from $\text{AP}_1$ with steering vector
 \begin{multline}
      \boldsymbol{a}(\theta,\phi )=[1, \cdots, \Omega^{M-1}, \Phi, \cdots,   \Omega^{M-1}\Phi,\cdots,  \Phi ^{N-1},\cdots, \\ \Omega^{M-1} \Phi ^{N-1}]^{T}
 \end{multline}

 Where $\Omega=e^{-j\pi \sin(\theta) \cos(\phi) }$,  $\Phi=e^{-j\pi  \sin(\theta) \sin(\phi)}$.
 For the roll angle $\gamma$ we use either the signal received from $\text{AP}_2$ or $\text{RIS}_1$ with the steering vector
 
 \begin{equation}
      \boldsymbol{a}(\gamma )=[1, \cdots, \Gamma^{M-1}, 1, \cdots,   \Gamma^{M-1},\cdots,  1,\cdots, \Gamma^{M-1} ]^{T}
 \end{equation}
 where $\Gamma=e^{-j\pi \sin(\gamma)}$.  In our environment, the two possible propagation paths that can be used for roll estimation have the same attenuation. We therefore do not specify which one is used. The \gls{rmse} of the orientation is evaluated numerically through simulation and turns out to be  $\sigma_0 = 0.4421$.

 As the beam training interval $T_{b}=102.4$~ms (e.g. one beacon interval in 802.11ad/ay) is large, and therefore the angle update rate $\frac{1}{T_{b}} \approx 9.8 \text{Hz} < B_m$ is lower than the movement bandwidth, we cannot directly use the \gls{aoa} estimate to determine the orientation. For this reason, we propose to re-calibrate the angle obtained by the gyroscope at every beam training. If we assume that the first beam training happens at time $0$, the gyroscope has an \gls{rmse} error of $\sigma_g$ and the error is identical and independent for all components, the uncertainty in the angle can be computed as\footnote{$u([a,b])$ is the characteristic function of the interval $[a,b]$}
 \begin{equation}
    \Vert \epsilon_a \Vert_{\text{RMS}} (t)  =  \sum_{k=0}^{\infty} \sqrt{\sigma^2_0 + 3 \sigma_g^2 T (t - k T_{b})} u([k T_{b}, (k+1) T_{b}]), \label{eq:gyro_and_aoa}
 \end{equation} 
 Whereas the uncertainty for the gyroscope is simply $\Vert \epsilon_a \Vert_{\text{RMS}} (t) = \left( \sqrt{3 T} \sigma_g\right)  \sqrt{t}$.
 Notably, the expression in equation \eqref{eq:gyro_and_aoa} is bounded in the interval $ [\sigma_0, \sqrt{\sigma^2_0 + 3 \sigma_g^2 T T_{b}}]$ for any $t$.

\section{Results}
\label{sec:results}

\begin{figure}[t]         
     \centering
    \input{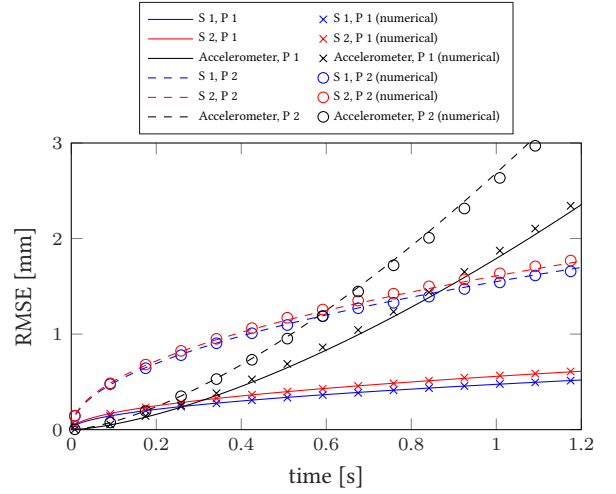}
    \vspace{-0.35cm}
     \caption{Positional error vs time.}  \vspace{-0.2cm}
     \label{fig:poserr_vs_time}
 \end{figure}

In this section, we compare the position and orientation estimates obtained from the \gls{jcs} method and the \gls{imu}. We consider two different propagation scenarios:
\begin{inparaenum}
    \item Scenario S1: 2 \gls{ap} and 1 \gls{ris}. $\text{AP}_1$ is used for the $x$ velocity component, as well as for azimuth and elevation, $\text{AP}_2$ for the $y$ velocity component and roll, and $\text{RIS}_1$ for the $z$ velocity component.  
    \item Scenario S2: 1  \gls{ap}  and 2 \glspl{ris}. $\text{AP}_1$ is used for the $x$ velocity component, as well as for azimuth and elevation, $\text{RIS}_2$ for the $y$ velocity component and roll, and $\text{RIS}_1$ for the $z$ velocity component.
\end{inparaenum}
As a typical off the shelf \gls{imu}, we use the device described in \cite{imu}, and we assume that the accelerometer is configured to the 4G and 16G acceleration ranges for the acceleration profiles P1 and P2 y. With this assumption, the noise variance $\sigma_a$ is $2$~mg for P1 and $3$~mg for P2.
Figure \ref{fig:poserr_vs_time} shows how the \gls{rms} of the position error of the \gls{jcs} and accelerometer evolves over time. In the plot, we can see how after around $0.25$ seconds for profile P1 and $0.6$ seconds for profile P2, the \gls{jcs} positional estimate outperforms the accelerometer. Moreover,  we also validated the theoretical formulation by numerically integrating 1000 realizations of the white noise evaluating the \gls{rmse}. As it can be seen in the plot, the theoretical \gls{rmse} matches with the numerical results.
 
 \begin{figure}[t]         
     \centering
    \input{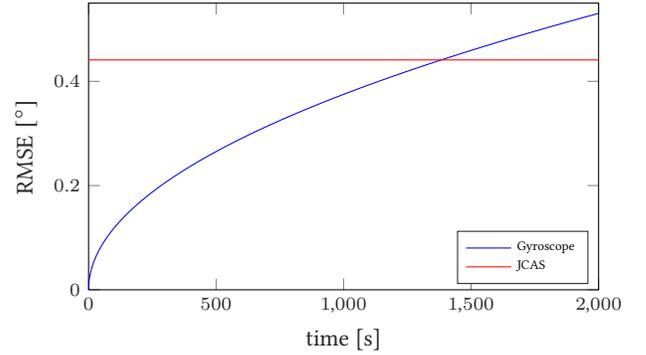}
    \vspace{-0.35cm}
     \caption{Angular error vs time.}  \vspace{-0.3cm}
     \label{fig:angerr_vs_time}
 \end{figure} 
In figure \ref{fig:angerr_vs_time} we show how the angular error evolves over time. In this case, the \gls{jcs} solution outperforms the \gls{imu} only after roughly 20 minutes, making its application impractical.

\section{Conclusion}
\label{sec:conclusion}

In this paper we show that, with the proposed scenario and model, integrating \gls{jcs}-based speed estimation can significantly improve \gls{hmd} tracking when visual/\gls{ir} tracking outage occurs. In particular, thanks to the fewer integration steps required, the estimate provided by the radio is more sustainable over long periods of time. The same can not be said for orientation though, where the classical \gls{imu} based tracking still outperforms \gls{jcs} for a long period of time. Although the numerical results are not representative of all possible scenarios, the fact that reducing the number of integration steps improves the sustainability of the tracking is. Further investigation, that we leave for future work, is needed to better quantify the gain in a more general setting. 

\begin{acks}
This work has received funding from the European Union’s EU Framework Programme for Research and Innovation Horizon 2020 under Grant Agreement No 861222.
\end{acks}
%%
%% The next two lines define the bibliography style to be used, and
%% the bibliography file.
\bibliographystyle{ACM-Reference-Format}
\bibliography{bibliography}

\end{document}